\documentclass[12pt,preprint,amsmath,amssymb,nofootinbib]{revtex4}
\usepackage{graphicx}

\begin{document}

\title{Could dark energy be vector-like?}

\author{C. Armend\'ariz-Pic\'on}

\affiliation{Enrico Fermi Institute and \\
  Department of  Astronomy and Astrophysics,\\
 University of Chicago. }

\begin{abstract}
  In this paper I explore whether  a vector field can be the origin of
  the  present  stage  of  cosmic  acceleration.  In  order  to  avoid
  violations  of  isotropy, the  vector  has  be  part of  a  ``cosmic
  triad'',  that is,  a set  of  three identical  vectors pointing  in
  mutually orthogonal  spatial directions. A  triad is indeed  able to
  drive a  stage of  late accelerated expansion  in the  universe, and
  there  exist  tracking   attractors  that  render  cosmic  evolution
  insensitive  to  initial  conditions.   However, as  in  most  other
  models,  the  onset  of  cosmic  acceleration  is  determined  by  a
  parameter that  has to be  tuned to reproduce current  observations. 
  The triad equation  of state can be sufficiently  close to minus one
  today, and for tachyonic models it  might be even less than that.  I
  briefly  analyze  linear  cosmological  perturbation theory  in  the
  presence  of  a   triad.   It  turns  out  that   the  existence  of
  non-vanishing spatial vectors invalidates the decomposition theorem,
  i.e.  scalar,  vector and tensor perturbations do  not decouple from
  each other.   In a  simplified case it  is possible  to analytically
  study the  stability of the  triad along the  different cosmological
  attractors.   The  triad  is  classically stable  during  inflation,
  radiation  and   matter  domination,  but  it   is  unstable  during
  (late-time) cosmic  acceleration.  I argue that  this instability is
  not likely to have a significant impact at present.
\end{abstract}

\maketitle

\section{Introduction}
A  combination  of different  cosmic  probes  that primarily  involves
supernova  data  \cite{Riess,  Knop}  suggests that  the  universe  is
presently  undergoing a  stage  of accelerated  expansion.  Little  is
known  about   the  origin  of  this  stage   of  cosmic  acceleration
\cite{Sean}.  It might be related to a breakdown of general relativity
on large  scales \cite{DGP, CDTT}, or  it can be the  effect of ``dark
energy'' \cite{darkenergy}, a  negative pressure component that causes
the  universe expansion  to accelerate.   The simplest  possibility is
that dark  energy merely is  a (tiny) cosmological constant.   If dark
energy  is dynamical,  it  is mostly  assumed  to be  a scalar  field,
quintessence \cite{quintessence}.

On  purely phenomenological  grounds,  within the  context of  general
relativity,  dark energy  might be  characterized by  its  equation of
state  $w_{DE}$,  its  speed  of  sound $c_s^2$  and  its  anisotropic
stresses     \cite{Wayne}.     Conventional     quintessence    models
\cite{quintessence} have an equation of state $-1\leq w_{DE}$, a speed
of sound  $c_s^2=1$ and no anisotropic stresses,  whereas in k-essence
models  \cite{kessence,Chiba}  the  speed   of  sound  is  in  general
different from one.  The range of possible phenomenological properties
of dark  energy is not exhausted  by the former  models.  Phantom dark
energy \cite{phantom}  has an equation of state  $w_{DE}<-1$, and thus
violates the dominant energy  condition \cite{CHT}. The origin of such
an equation of state is the ``wrong'' sign kinetic term of the phantom
scalar   field.    Because   of   that,   phantom   dark   energy   is
quantum-mechanically unstable upon decay  of the vacuum into (positive
energy)   gravitons   and    (negative   energy)   phantom   particles
\cite{CHT,CJM}.   Hence, if future  preciser observations  confirm the
current trend and  favor a dark energy equation  of state smaller than
minus  one \cite{Riess,Starobinsky},  alternative (viable)  models are
needed  to   account  for  such   a  value\footnote{The  vacuum-driven
  metamorphosis  model of Parker  \& Raval  \cite{Parker} seems  to be
  experimentally   ruled  out   at  the   $99.5\%$   confidence  level
  \cite{Riess}.}  \cite{MMOT}.

In  this paper  I  explore whether  dark  energy can  be vector-like.  
Vector-like dark  energy turns out  to display a series  of properties
that  make  it particularly  interesting  phenomenologically.  On  one
side, it can also  violate the dominant energy condition, $w_{DE}<-1$,
while possessing a  conventional kinetic term.  On the  other side, it
has non-anisotropic stress perturbations and it leads to violations of
the decomposition theorem \cite{KodamaSasaki}, i.e.  the decoupling of
scalar,  vector and  tensor cosmological  perturbations.  As  we shall
see,  other interesting  features  include the  existence of  tracking
attractors,  that  render  the  evolution  of a  vector  field  in  an
expanding universe rather insensitive to initial conditions.  In spite
of  this  attractive feature,  the  onset  of  cosmic acceleration  is
determined by  a parameter in the  Lagrangian that has  to be properly
adjusted, as  in most other models.   In that respect,  these forms of
vector   dark   energy    are   similar   to   quintessence   trackers
\cite{tracker,Chiba}.

Non-gravitational  interactions are  known  to be  mediated by  vector
fields.  In  addition, from a four-dimensional point  of view, certain
components   of  higher-dimensional   metrics   behave  like   vectors
\cite{KaluzaKlein}.   It is hence  natural to  study the  evolution of
vector fields in  a cosmological setting. However, the  existence of a
spatially   non-vanishing   vector    breaks   the   isotropy   of   a
Friedmann-Robertson-Walker (FRW) universe.  From  the point of view of
gravity,  such a  breaking  manifests itself  in anisotropic  stresses
caused by the vector.  If dark energy  is such a vector, as long as it
remains subdominant,  this violation  is likely to  be observationally
irrelevant  \cite{stresses}.   Once  dark  energy  comes  to  dominate
though, one would expect an  anisotropic expansion of the universe, in
conflict  with  the  significant  isotropy  of  the  cosmic  microwave
background (CMB)  \cite{isotropy}.  Because of  that, in this  paper I
consider a ``cosmic triad'', i.e.  a set of three equal length vectors
that  point   in  three   mutually  orthogonal  spatial   directions.  
Remarkably,  the existence  of a  triad  turns to  be compatible  with
spatial isotropy, at  least from the point of  view of gravity.  While
the  triad guarantees  the isotropy  of  the background,  it does  not
automatically imply the isotropy of its perturbations.  Eventually, it
might be  even necessary to  introduce fields that  explicitly violate
rotational  symmetry, as  there appear  to be  hints  of (statistical)
anisotropy in  the CMB fluctuations  \cite{non-isotropy}.  Along these
lines, I  speculate below  that a triad  could provide a  link between
cosmic  acceleration and  some of  the anomalies  observed in  the CMB
radiation \cite{non-isotropy,quadrupole}.

Mainly because they single  out spatial directions, vector fields have
received  comparatively  little  attention  in  cosmology.   Ford  has
proposed an  inflationary model  where a vector  is responsible  for a
stage of inflation \cite{Ford}.  Our  treatment here is to some extent
similar  to his  proposal.  Jacobson  and Mattingly  have  studied the
dynamics of a  vector with of fixed length,  with the specific purpose
of studying violations of Lorentz invariance \cite{JacobsonMattingly}.
A vector-like form of quintessence has been also considered by Kiselev
\cite{Kiselev},  though his  vectors significantly  differ from  ours. 
Zimdahl {\it  et al.}  have  suggested that a (timelike)  vector force
could  be responsible  for the  present acceleration  of  the universe
\cite{Zimdahl}.   Also,  it  has  been  noted  that  the  addition  of
higher-order powers of the Maxwell field-strength to the Lagrangian of
an  electromagnetic  field  might  cause the  universe  to  accelerate
\cite{NPS}.  The  literature on magnetic fields in  the early universe
is more extensive, see \cite{magnetic} and references therein.

\section{Vector dark energy}
Consider  a set  of three  self-interacting vector  fields  $A^a_\mu$. 
Strictly  speaking, this is  really a  set of  three one-forms,  but I
shall  call them vectors.   Latin indices  label the  different fields
($a, b, \ldots=1\ldots 3$) and greek indices their different spacetime
components ($\mu,\nu,\ldots =0\ldots 3$).  As we shall see below, this
number  of  vector  fields  is  dictated  by  the  number  of  spatial
dimensions and  the requirement of  isotropy.  We would like  to study
the dynamics  of such a ``cosmic  triad'' in the presence  of gravity. 
Consider hence those vectors minimally coupled to general relativity,
\begin{equation}\label{eq:action}
  S=\int d^4x \sqrt{-g}\left[\frac{R}{16\pi G}
    -\sum_{a=1}^{3}\left(\frac{1}{4}F^a_{\mu\nu}F^a{}^{\mu\nu}
      +V(A^a{}^2)\right)+
    \mathcal{L}_m(g_{\mu\nu}, \psi)\right],
\end{equation}
where  $F_{\mu\nu}^a=\partial_\mu  A^a_\nu-\partial_\nu  A^a_\mu$  and
$A^a{}^2=g^{\mu\nu}A^a_\mu  A^a_\nu$.   The  action  (\ref{eq:action})
thus contains  three identical  copies of the  Lagrangian of  a single
vector  field.   The  term  $V(A^a{}^2)$ is  a  self-interaction  that
explicitly  violates  gauge  invariance.   For  completeness,  in  the
Appendix I show how a triad could naturally appear from a gauge theory
with  a single  $SU(2)$  gauge group.   In  the following,  I use  the
Einstein  summation  convention throughout,  where  a  sum is  implied
\emph{only}  over indices  in  opposite positions.   The indices  that
label  the different  vectors are  raised  and lowered  with the  flat
``metric'' $\mathcal \delta_{ab}$.

The kinetic term  $F^2$ in the action (\ref{eq:action})  is not unique
in   the  following   sense.   Up   to  boundary   terms,  $\nabla_\nu
A_\mu\nabla^\nu A^\mu$ is the only additional diffeomorphism invariant
quadratic  term that contains  two derivatives  of the  vector $A_\mu$
\cite{Will,JacobsonMattingly}.  Because the  dynamics of vectors known
to occur in nature are described by a Maxwell term, I consider a $F^2$
term  only.  Additional  couplings of  the vector  are  constrained by
tests  of gravity  \cite{Will} and  limits on  possible  violations of
Lorentz symmetry\footnote{Note that in a FRW universe Lorentz symmetry
  is  (spontaneously)  broken anyway,  in  the  sense  that there  are
  non-vanishing vector  fields, like the gradient of  the Ricci scalar
  or the CMB  temperature, that define a preferred  direction.  Such a
  breaking could  be detected by non-gravitational  experiments if the
  non-vanishing vector directly  couples to matter \cite{Lehnert}. See
  also  \cite{covariance}  for  a  clear discussion  of  the  relation
  between  coordinate invariance,  Lorentz invariance  and isotropy.}  
\cite{Lorentz}.  To  avoid such violations,  I assume that  the matter
Lagrangian $\mathcal{L}_m$ only depends on the metric $g_{\mu\nu}$ and
on the  remaining matter  fields $\psi$, but  not on the  cosmic triad
$A^a$.   In  that respect,  the  triad  is  analogous to  conventional
quintessence \cite{rest}.

Varying the  action (\ref{eq:action}) with  respect to the  metric one
obtains the  Einstein equations $G_{\mu\nu}=8\pi  G T_{\mu\nu}$, where
the energy momentum tensor of the triad is given by
\begin{equation}\label{eq:EMT}
  {}^{(A)}T_{\mu\nu}=\sum_a \left[F^a_{\mu\rho}F^a_\nu{}^\rho
  +2 \frac{dV}{dA^a{}^2}A^a_\mu A^a_\nu
  -\left(\frac{1}{4}F^a_{\rho\sigma}F^a{}^{\rho\sigma}
    +V(A^a{}^2)\right)g_{\mu\nu}\right].
\end{equation}
This energy momentum  tensor is the sum of  the three different energy
momentum tensors of  the decoupled vectors, ${}^{(A)}T_{\mu\nu}=\sum_a
{}^{(a)}T_{\mu\nu}$, neither  of which  is of perfect-fluid  form.  By
varying the action with respect  to the vectors $A^a_\mu$, one obtains
their equations of motion,
\begin{equation}\label{eq:fullmotion}
  \partial_\mu (\sqrt{-g} F^a{}^{\mu\nu})
  =2 \sqrt{-g} \frac{dV}{dA^a{}^2} A^a{}^\nu.
\end{equation}
The four-divergence  of the last equation yields  a constrain equation
for the  vector.  As  a consequence, each  vector has  three dynamical
degrees of freedom, as it should.
 
We shall study the dynamics of these vectors in a flat, homogeneous
and isotropic FRW universe with metric
\begin{equation}\label{eq:FRW}
  ds^2=-dt^2+a^2(t)d\vec{x}^2.
\end{equation}
An  ansatz  for the  vectors  that turns  to  be  compatible with  the
symmetries of this metric (homogeneity and isotropy) is
\begin{equation}\label{eq:ansatz}
  A^b{}_\mu=\delta^b{}_\mu \, A(t)\cdot a.
\end{equation}
Hence  the three vectors  point in  three mutually  orthogonal spatial
directions   and   they   share   the  same   time-dependent   length,
$A^a{}^2\equiv  A^a_\mu  A^{a\mu}=A^2(t)$.   Substituting  the  ansatz
(\ref{eq:ansatz})  and  the  metric  (\ref{eq:FRW})  into  the  vector
equations of motion (\ref{eq:fullmotion}) I find
\begin{equation}\label{eq:xmotion}
  \ddot{A}+3H\dot{A}+\left(H^2+\frac{\ddot{a}}{a}\right)A+
\frac{dV}{dA}=0,
\end{equation}
where a dot means a derivative  with respect to cosmic time $t$.  Note
that  the  $0$-component   of  equation  (\ref{eq:fullmotion})  forces
$A^a_0$ to  vanish, as  in the ansatz  (\ref{eq:ansatz}). Substituting
the metric (\ref{eq:FRW}) into the Einstein equations one obtains
\begin{subequations} \label{eq:Einstein}
  \begin{eqnarray}
    H^2&=&\frac{8\pi G}{3}\rho, \label{eq:Friedmann}\\
    \frac{\ddot{a}}{a}&=&-\frac{4\pi G}{3}(\rho+3p),
  \end{eqnarray}
\end{subequations}
where  $H\equiv \dot{a}/a$  is  the Hubble  ``constant''.  The  energy
density of the  universe is $\rho\equiv -T^0{}_0$ and  its pressure is
defined by $p  \cdot \delta^i_j\equiv T^i{}_j$, where $i$  and $j$ run
over the spatial spacetime  components.  Note that the energy momentum
tensor has  to be compatible with  the symmetries of  the metric.  For
the FRW metric (\ref{eq:FRW}), $G^0{}_i=0$, so that $T{}^0{}_i$ should
also vanish.  It  can be easily verified that this  is indeed the case
for  the ansatz  (\ref{eq:ansatz}).   The requirement  of isotropy  is
non-trivial for a single vector, since its energy momentum tensor is
\begin{equation}
  {}^{(a)}T^i{}_j=\left(\frac{1}{2}(\dot{A}+H A)^2-V\right)
  \delta^i{}_j
  +\left(2\frac{dV}{dA^2}A^2-(\dot{A}+H A)^2\right)
  \delta^i{}_j \cdot \delta^a{}_j. 
\end{equation}
Although this energy momentum tensor  is diagonal, its value along the
$j=a$  direction  is  different  from  the one  along  the  directions
perpendicular to  it.  Nevertheless, the total  energy momentum tensor
${}^{(A)}T_{\mu\nu}\equiv  \sum_a  {}^{(a)}T_{\mu\nu}$  has  isotropic
stresses,   and  the   corresponding  energy   density   $\rho_A$  and
(isotropic) pressure $p_A$ are given by
\begin{subequations}\label{eq:pandrho}
\begin{eqnarray}
  \rho_A&=&\frac{3}{2}(\dot{A}+H A)^2+3V(A^2), \label{eq:rho}\\
  p_A&=&\frac{1}{2}(\dot{A}+H A)^2-3V(A^2) \label{eq:p}
  +2\frac{dV}{dA^2}A^2.
\end{eqnarray}
\end{subequations}
Note  that  the equation  of  motion  (\ref{eq:xmotion})  can be  also
derived from the condition ${\dot{\rho}_A+3H (\rho_A+p_A)=0}$.

To conclude this section let me point out a remarkable property of the
cosmic triad. Namely, its equation of state $w_A\equiv p_A/\rho_A$ can
become less than $-1$ (with a positive energy density) if $dV/dA^2$ is
negative \cite{AD}.  Because a mass  term for a vector $A_\mu$ has the
form  $V(A^2)=  \frac{m^2}{2}  A_\mu   A^\mu$,  I  call  such  vectors
``tachyonic''.   Tachyons  (particles of  negative  squared mass)  are
usually associated with instabilities.   In many cases, an instability
merely signals  the tendency of  the system to evolve.   In cosmology,
those  instabilities  are  not  particularly  terrible.   In  fact,  a
universe in stable  equilibrium would be pretty lame,  as it would not
even expand.   In the absence of  gravitational instability structures
would  not form,  and  without  the (effective)  tachyonic  mass of  a
scalar, it would be quite difficult to seed a scale invariant spectrum
of  perturbations during inflation  \cite{Mukhanov,liberated}.  Scalar
tachyons\footnote{By  a  ``scalar  tachyon''  I mean  a  scalar  field
  $\varphi$  with a  convex potential,  $d^2  V/d\varphi^2<0$.} indeed
have  been  widely  considered  in  the literature.   Other  forms  of
instability   are   more   worrisome,  like   the   quantum-mechanical
instability of the vacuum in the presence of a phantom \cite{CHT,CJM}.
By simple analogy, any form of instability associated with a tachyonic
vector is not expected to be  of this second kind, as the vectors have
a conventional  kinetic term.  This  question is not only  of academic
interest, since  analyses of  observational data (marginally)  favor a
dark energy equation of state $w_{DE}<-1$ \cite{Riess,Starobinsky}.

I  shall  not  deal  here  with the  quantum  mechanics  of  tachyonic
particles,  which  even   for  scalars  is  not  free   of  problems.  
Nevertheless, I also want to  present some arguments that suggest that
a   tachyonic   vector  might   be   similar   to   a  phantom   field
\cite{Dubovsky}.  One  of the arguments goes back  to the Stueckelberg
theory  of  massive   vectors  \cite{Stueckelberg,RR}.   Consider  for
simplicity a vector field in Minkowski space,
\begin{equation}\label{eq:massive} 
 L=-\frac{1}{4}F_{\mu\nu}F^{\mu\nu}-\frac{\kappa m^2}{2}A_\mu A^\mu.
\end{equation}
The field  is massive  for $\kappa=1$ and  tachyonic for  $\kappa=-1$. 
Upon  the substitution  ${A_\mu\to  A_\mu+m^{-1}\partial_\mu S}$,  the
Lagrangian (\ref{eq:massive}) reads
\begin{equation}\label{eq:Stueckelberg} 
  L=-\frac{1}{4}F_{\mu\nu}F^{\mu\nu}-\frac{\kappa m^2}{2}A_\mu A^\mu
  -\kappa m A^\mu \partial_\mu S
  -\frac{\kappa}{2}\partial_\mu S\partial^\mu S.
\end{equation}
Note that  (\ref{eq:Stueckelberg}) contains an  additional scalar, the
Stueckelberg field  $S$. For a  massive vector ($\kappa=1$) $S$  has a
conventional kinetic  term, but for a  tachyonic vector ($\kappa=-1$),
its kinetic  term has the  ``wrong'' sign, like  the one of  a phantom
\cite{phantom}.   However,  the  additional  field  $S$  turns  to  be
constrained in the quantum theory.  Even though it describes a massive
vector, the  Lagrangian (\ref{eq:Stueckelberg}) has  a gauge symmetry,
${A_\mu\to A_\mu+\partial_\mu  \lambda}$, ${S \to  S-m\lambda}$, where
$\lambda$ is a scalar function.  Upon quantization, this gauge freedom
is   fixed   by  imposing   the   Stueckelberg  subsidiary   condition
${\partial_\mu A^\mu-m S=0}$, which  relates $S$ and the divergence of
the vector \cite{RR}.   Thus, strictly speaking, the field  $S$ is not
simply a phantom scalar in the conventional sense.  On the other hand,
there  are other properties  that suggest  phantom-like behavior  of a
tachyonic  vector, like  the opposite  sign of  the propagator  in the
high-momentum  limit, or  the opposite  sign in  front of  the squared
longitudinal momentum in the Hamiltonian \cite{Weinberg}.

In the time-dependent  situation we are dealing with,  where the triad
vectors have a non-vanishing  expectation value, the issue is slightly
more complicated.  Consider quantum fluctuations $\delta A_\mu$ around
one  of the  triad vectors  in our  classical  solutions, ${A^a_\mu\to
  A^a_\mu  +\delta  A_\mu}$.    Expanding  the  vector  Lagrangian  in
(\ref{eq:action})  to second  order in  $\delta A_\mu$  and neglecting
fluctuations in  the gravitational field I  get, for one  of the triad
vectors,
\begin{equation}
  L=-\frac{1}{4}\delta F_{\mu\nu}\delta F^{\mu\nu}
  -\frac{dV}{dA^2}\delta A_\mu \delta A^\mu
  -2\frac{d^2 V}{d^2 A^2} (A^{a\mu} \delta A_\mu)^2.
\end{equation}
Terms linear  in the perturbations vanish if  $A^{a\mu}$ satisfies the
classical equations of motion. Note  that in addition to the mass term
proportional  to  $dV/dA^2$,   there  is  an  additional  contribution
proportional to $d^2 V/d^2A^2$ that breaks Lorentz invariance (because
of  the coupling  of $\delta  A_\mu$ to  the  classical, non-vanishing
vector  $A^\mu$.)  Therefore,  the quantization  of $\delta  A_\mu$ is
expected to be  significantly different from the one  of the tachyonic
vector in the Lagrangian (\ref{eq:massive}).

\section{Cosmic evolution}\label{sec:evolution}

Our next task  consists in studying the evolution  of the cosmic triad
in a  universe that contains  additional forms of ``matter'',  like an
inflaton\footnote{The   inflaton  is   the  component   or  components
  responsible  for an eventual  stage of  inflation.} ($p_{inf}\approx
-\rho_{inf}$), radiation ($p_r=\rho_r/3$) or dust ($p_d=0$).  Ideally,
we would  like the cosmic triad  to remain subdominant  during most of
cosmic history, and just around redshift $z\approx 1$ come to dominate
the energy density of the  universe and trigger a stage of accelerated
expansion.

The vector equation of  motion (\ref{eq:xmotion}) is formally the same
as the  one for  a self-interacting \emph{conformally  coupled} scalar
field.  Indeed, the  term in parenthesis is proportional  to the Ricci
scalar $R$,  which vanishes  during radiation domination.   Of course,
the  similarity between  the equations  of motion  of a  vector  and a
conformally coupled scalar arises from the conformal invariance of the
Maxwell  Lagrangian.   In  some  instances  it is  going  to  be  more
convenient  to  deal  with  a  set of  two  first  order  differential
equations, rather  that with a  single second order  one.  Introducing
the number  of $e$-folds  $dN\equiv d\log a$  as a time  variable, and
defining
\begin{equation}
  B=\dot{A}+H A,
\end{equation}
the vector equation of motion (\ref{eq:xmotion}) can be recast as
\begin{subequations}\label{eq:xandy}
\begin{eqnarray}
  \frac{dB}{dN}+2 B +\frac{1}{H}\frac{dV}{dA}&=&0, \\
  \frac{dA}{dN}-\frac{B}{H}+A&=&0 \label{eq:onlyx},
\end{eqnarray}
\end{subequations}
where the Hubble constant is given by equation (\ref{eq:Friedmann}).

In the following I study the vector equations of motion in two limits:
the limit  where matter dominates  (early stages of  cosmic evolution)
and the limit of triad domination (late stages).  The evolution of the
triad depends on the self interaction  $V$. In this paper I focus on a
class of run-away power-law potentials
\begin{equation}\label{eq:V}
  V(A^2)=M^4\cdot \left(\frac{A^2}{M^2}\right)^{-n},
\end{equation}
where $M$ is a positive parameter with dimensions of mass.  This class
turns to be sufficiently general  for our purposes.  In five spacetime
dimensions, a  self-interaction of this  form (with $n=1/2$)  leads to
inflation along  ``our'' three  spatial dimensions, while  keeping the
size of the remaining  fifth dimension essentially constant \cite{AD}. 
These  runaway   potentials  are  also  reminiscent  of   a  class  of
``tracker'' quintessence  \cite{tracker,Chiba}.  Note that  for $n>0$,
the  case  we are  interested  in, all  these  models  are tachyonic.  
Non-tachyonic   interactions  do   not  appear   to   be  particularly
interesting.

\subsection{Matter domination}
\begin{figure}
  \includegraphics{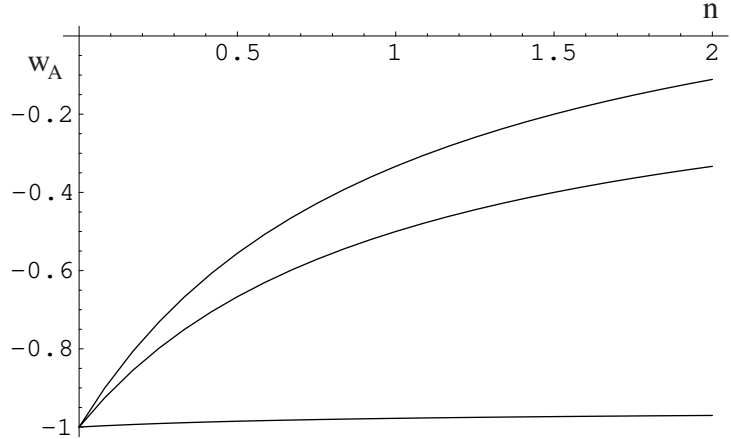}
    \caption{A plot of $w_A$, equation (\ref{eq:wattractor}), 
      for  (from top  to bottom)  $\beta=1/2$  (radiation domination),
      $\beta=2/3$ (dust domination) and $\beta=15$ (nearly de Sitter
      inflation).
      \label{fig:wattractor}}
\end{figure}

Suppose that  the energy density of  the universe is  dominated by the
inflaton, radiation  or dust, i.e.   the contribution of the  triad to
the total  energy density  of the universe  is negligible.   The scale
factor then grows like
\begin{equation}\label{eq:a}
  a(t)=a_0 \left(\frac{t}{t_0}\right)^\beta, \quad 
\text{where} \quad \beta=\frac{2}{3(1+w_m)}
\end{equation}
and $w_m$ is the matter  equation of state.  Thus, $\beta\gg 1$ during
nearly  de Sitter inflation,  $\beta=1/2$ during  radiation domination
and $\beta=2/3$  during dust domination.   We shall keep $\beta$  as a
free parameter and consider solutions  of the equation of motion where
$A$ also grows as a power  in cosmic time.  The reader can verify that
indeed,
\begin{equation}\label{eq:mattersolution}
  A=\left(\frac{2n (n+1)^2}{3\beta(n+1)
      -n+\beta(2\beta-1)(n+1)^2}\right)^{\frac{1}{2n+2}}
  \cdot M 
  \cdot\left(\frac{t}{M^{-1}}\right)^{\frac{1}{1+n}}
\end{equation}
is  a solution  of the  equation of  motion (\ref{eq:xmotion})  for an
interaction given  by equation (\ref{eq:V}).   Actually, this solution
is also  an attractor.  Perturbing  $A\to A+\delta A$  and linearizing
equation  (\ref{eq:xmotion})  for  the  given  unperturbed  background
(\ref{eq:a}) I find
\begin{equation}
  \delta\ddot{A}+3H \delta\dot{A}
  +\left(H^2+\frac{\ddot{a}}{a}\right)\delta A
  +\frac{d^2V}{dA^2}\delta A=0.
\end{equation}
The term  in parenthesis  is positive during  dust domination,  and it
vanishes  during   radiation  domination.   At  the   same  time,  the
derivative  $d^2  V/dA^2$  is   always  positive  for  the  potentials
(\ref{eq:V}).    Therefore,  small   deviations   from  the   solution
(\ref{eq:mattersolution})  oscillate and decay  away. Note  that along
the attractor, the equation of state of the triad is constant,
\begin{equation}\label{eq:wattractor}
  1+w_A=\frac{n}{1+n}(1+w_m).
\end{equation}
Hence, these solutions are  analogous to the quintessence and trackers
discussed in  \cite{tracker} and \cite{Chiba}.  The  triad equation of
state is  always smaller than the  one of the  dominant component, and
the closer  the latter  to $-1$, the  lesser is their  difference. The
triad equation  of state as a  function of $n$ is  plotted for various
values  of $\beta$  in  figure \ref{fig:wattractor}.   Because at  the
present  epoch the universe  is not  dominated by  dust any  more, the
value of the triad equation of state today lies between the ones along
the matter attractor and the de  Sitter attractor I will discuss next. 
Hence,  experimental  constraints \cite{Riess,Knop}  on  the value  of
$w_{DE}$ restrict the possible values of $n$.  In particular, in order
to obtain $w_A\lessapprox 0.75$  today, models with $n\lessapprox 1/2$
have  to be  considered.   This restriction  applies provided  initial
conditions  are chosen  in the  basin  of attraction  of the  tracking
solution.

For later purposes, let me also discuss an approximate solution of the
system (\ref{eq:xandy}).  Suppose that $B \ll H A$ and $dV/dA\ll H B$.
Loosely  speaking, these inequalities  are attained  in the  limits of
large  $A$ or large  $H$.  Then,  equations (\ref{eq:xandy})  have the
approximate solution
\begin{equation}\label{eq:approximate}
  A=A_{(0)} \exp[-(N-N_0)], \quad B=B_{(0)} \exp[-2(N-N_0)]
\end{equation}
Along  this  solution the  kinetic  energy  of  the field  $B^2\propto
e^{-4N}$ decreases,  whereas the potential energy  $V\propto e^{2n N}$
increases. Hence, soon the potential energy dominates the kinetic one,
$B^2\ll V$,  so that along  this approximate solution the  equation of
state is
\begin{equation}\label{eq:wapproximate}
  w_A\approx-1-\frac{2}{3}n,
\end{equation} 
which  is less  than $-1$.   However, this  approximate  solution only
holds  temporarily, as  the assumptions  we have  made  finally become
violated.  Later on I will discuss an example where initial conditions
place the triad on the approximate solution (\ref{eq:approximate}) for
a significant  period of  time. Note that  in the opposite  limit, the
limit where  the kinetic energy  dominates the potential  one, $B^2\gg
V$, the triad  equation of state is radiation-like,  $w_A\approx 1/3$. 
In both limits, the triad is quite different from a (canonical) scalar
field (with positive potential), for which $-1\leq w\leq 1$.

\subsection{Triad domination}
Along the  attractor (\ref{eq:mattersolution}), the  energy density of
the triad decays  slower than the one of  matter. Therefore, sooner or
later the triad will come  to dominate the universe. Consider thus the
equation  of motion  (\ref{eq:xmotion}) when  the triad  dominates the
content  of  the universe.   Using  equations (\ref{eq:Einstein})  and
(\ref{eq:pandrho})  the vector  equation of  motion (\ref{eq:xmotion})
takes the form
\begin{equation}\label{eq:dominantx}
\ddot{A}+3H\dot{A}+\frac{dW}{dA}=0, \quad \mathrm{where} \quad
\frac{dW}{dA}\equiv \frac{4\pi G}{3}\left(12 V- 3\frac{dV}{dA}A\right)A
  +\frac{dV}{dA},
\end{equation}
which  is just  like the  equation of  motion of  a  minimally coupled
scalar with appropriate potential (we  won't need the explicit form of
$W(A)$).

\begin{figure}
  \includegraphics{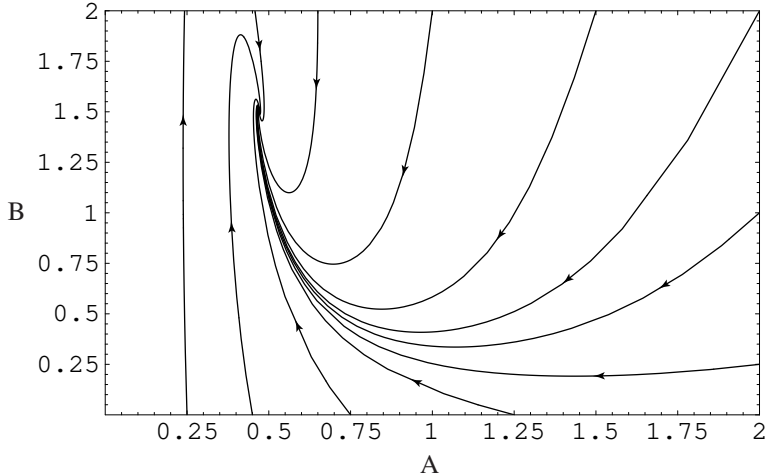}
    \caption{A phase diagram of the system (\ref{eq:xandy}) for $n=1/2$.
      The  units are arbitrary.   The diagram  clearly shows  that all
      phase trajectories converge to the single de Sitter attractor at
      constant $A$ and $H$.
      \label{fig:phase}}
\end{figure}

Solutions of  equation (\ref{eq:dominantx}) with  constant $A=A_*$ can
be easily  identified by requiring  $dW/dA$ to have  a zero at  $A_*$. 
This leads to the condition
\begin{equation}\label{eq:dSA}
  \frac{d\log V}{d\log(4\pi G A^2-1)}\bigg|_{A_*}=2.
\end{equation} 
Those constant $A$ solutions are expected to be stable (attractors) if
$d^2W/dA^2>0$, which implies
\begin{equation}\label{eq:stability}
  \frac{dV}{dA^2}+\frac{1}{4\pi G}\frac{d^2V}{d^2A^2}(1-4 \pi G A^2)
  \bigg|_{A_*}>0.
\end{equation}
Along   the  attractor   the   energy  density   is,  from   equations
(\ref{eq:Friedmann}) and (\ref{eq:rho}),
\begin{equation}\label{eq:density}
  \rho_A=\frac{3 V(A_*^2)}{1-4\pi G \, A_*^2},
\end{equation}
i.e. a  constant. Hence,  these constant $A$  solutions are  de Sitter
attractors with ${w_A=-1}$; they are natural candidates to accommodate
the   present   accelerated  expansion   of   the  universe.    Figure
\ref{fig:phase}    shows    a   phase    diagram    of   the    system
(\ref{eq:dominantx}) when  the triad is the dominant  component of the
universe. Note that along the  de Sitter attractor, the kinetic energy
of the triad does not vanish.

Inserting the potentials  (\ref{eq:V}) into equation (\ref{eq:dSA})
and  verifying condition (\ref{eq:stability})  I find  that there  is a
single (stable) de Sitter attractor at
\begin{equation}\label{eq:dS}
  A_*^2=\frac{1}{4\pi G}\frac{n}{n+2}.
\end{equation}
Substituting this value of $A_*$ into (\ref{eq:density}) and requiring
it to  be of  the order of  the present  energy density, one  can then
estimate what is the required value of $M$ to fit the present stage of
accelerated expansion,
\begin{equation} 
  M\sim 10^{-\frac{122}{4+2n}}G^{-1/2}.
\end{equation}
Therefore,  for   $n\approx  0$  the  required  value   leads  to  the
``infamous'' scale  $M\sim 10^{-3}$ eV,  whereas larger values  of $n$
result  into more  reasonable  energies.  Note  that  we are  fitting,
rather  that  explaining, the  time  cosmic  acceleration begins.   As
mentioned above, the exponent $n$ determines the value of the equation
of state  today. The  close $n$ to  $0$, the  closer $w_A$ is  to $-1$
today.

\subsection{Two Examples}

In  this  section  I  present  two  particular  examples  of  possible
realizations of vector dark energy. Current limits on the value of the
equation  of state  of dark  energy $w_{DE}$  and its  derivative with
respect to redshift $dw_{DE}/dz$ at $z=0$ are
\begin{equation}\label{eq:limits}
  w_{DE}\big|_{z=0}=-1.31\pm^{0.22}_{0.28} \quad \text{and} \quad
  \frac{dw_{DE}}{dz}\Big|_{z=0}=1.48\pm^{0.81}_{0.90},
\end{equation}
where    the    prior   on    the    density    parameter   of    dust
$\Omega_M=0.27\pm0.04$  has been  assumed  \cite{Riess}.  Because  the
uncertainties are  correlated, the  reader is advised  to look  at the
constraints  on   the  $w_{DE}-dw_{DE}/dz$  plane  in   figure  10  of
\cite{Riess}.    Note   that    the   limits   (\ref{eq:limits})   are
significantly weaker  than the ones derived assuming  that $w_{DE}$ is
constant.

\begin{figure}
  \includegraphics{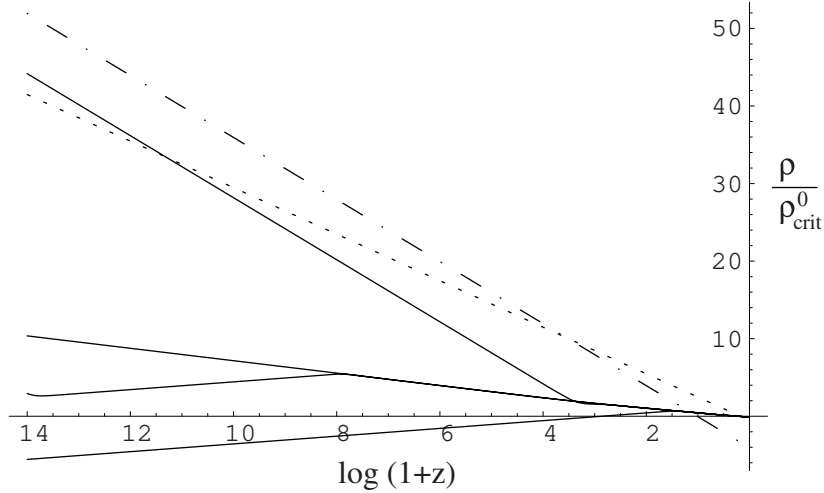}
    \caption{Triad energy density (in units
      of  today's critical energy  density) versus  $\log_{10} (1+z)$
      for  the  interaction  (\ref{eq:interaction1}).   Shown  is  the
      energy density  of vector dark  energy for four sets  of initial
      conditions  (continuous  lines).    For  reference,  the  energy
      densities of radiation (dash-dotted)  and dust (dotted) are also
      displayed.  Note that despite  the big difference in the initial
      value  of  the energy  density  (50  orders  of magnitude),  the
      present  value of  the dark  energy density  parameter  still is
      $\Omega_{DE}\approx 0.72$.
      \label{fig:rho}}
\end{figure}

\begin{figure}
  \includegraphics{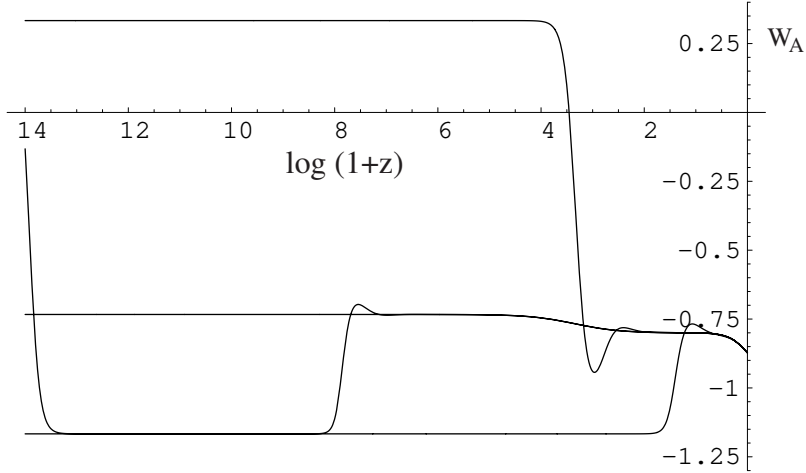}
    \caption{A plot of the triad equation of state 
      for  the same initial  conditions and  interaction as  in figure
      \ref{fig:rho}.   The attractors are  easily identified  by their
      constant  value   of  $w_A$,  which  can  be   read  off  figure
      \ref{fig:wattractor}.   For one set  of initial  conditions, the
      system reaches  first the radiation attractor,  then proceeds to
      the dust attractor and finally  to the de Sitter attractor.  For
      another, the system does  not reach the radiation attractor, but
      does  reach the  dust one  before  continuing to  the de  Sitter
      solution.   And lastly,  for yet  a different  one  the universe
      barely reaches the transition to the de Sitter attractor just in
      time.  For all initial conditions, $w_{A}\approx -0.87$ today.
      \label{fig:w}}
\end{figure}

The first example has a potential
\begin{equation}\label{eq:interaction1}
  V(A^2)= M^4 \left(\frac{A^2}{M^2}\right)^{-1/4}, \quad \textrm{where}
  \quad M^{9/2}= 9\cdot 10^{-2} \left(\frac{3}{8\pi G}\right)^{5/4} 
  H_0^2
\end{equation}
and I am assuming that the Hubble constant today is $H_0=70$ km/s/Mpc.
Note that in  this model $V$ is not an analytic  function of $A^2$. In
particular,   $A^2$   is   constrained   to  be   spacelike.    Figure
\ref{fig:rho} shows the evolution  of the energy density for different
sets  of   initial  conditions,  and  figure   \ref{fig:w}  shows  the
corresponding  behavior of the  triad equation  of state.   As clearly
seen in  the figures, the late  time evolution of the  cosmic triad is
quite insensitive  to initial conditions.  For  this particular model,
$w_A\approx -0.87$,  $dw_A/dz\approx 0.10$ and  $\Omega_A\approx 0.72$
today.  These  values are  consistent with the  current limits  on the
properties  of dark energy  (\ref{eq:limits}), see  also figure  10 in
\cite{Riess}.  Larger values of $n$ lead to violations of the limit on
$w_{DE}$ today, whereas smaller values yield $w_A$'s closer to $-1$.

\begin{figure}
  \includegraphics{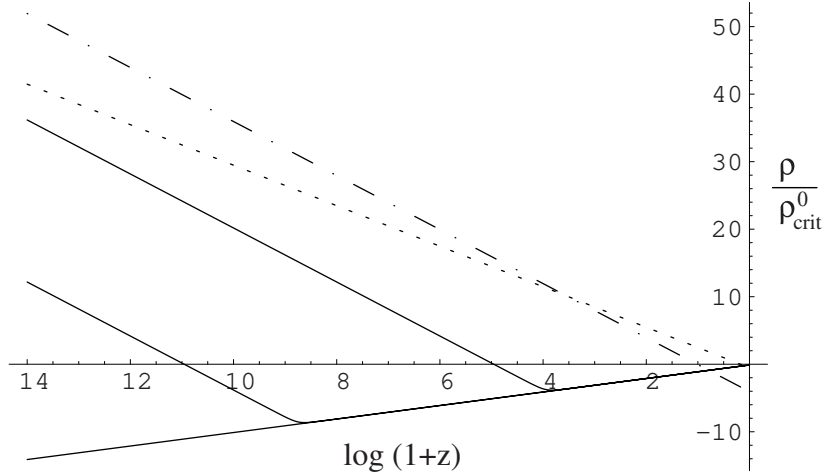}
    \caption{Triad energy density (in units
      of today's critical energy density) versus $\log_{10} (1+z)$ for
      the  second model  (\ref{eq:interaction2}).  For  reference, the
      energy  densities of radiation  (dash-dotted) and  dust (dotted)
      are  also  displayed.  The  shown  trajectories  share the  same
      initial  value of  $A$, though  the  initial values  of $B$  are
      different.   The  present  value  of  the  dark  energy  density
      parameter is $\Omega_{DE}\approx 0.72$.
      \label{fig:rho2}}
\end{figure}

\begin{figure}
  \includegraphics{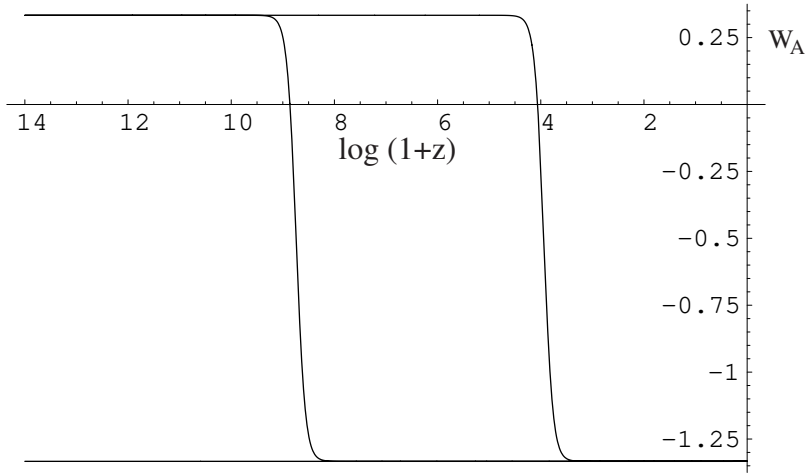}
    \caption{A plot of the triad equation of state of the dark 
      triad  for different initial conditions (same  as in Figure
      \ref{fig:rho2}).  For all of them the triad reaches the equation
      of  state   $w_A=-4/3$,  equation  (\ref{eq:wapproximate}),  and
      remains at that value till the present epoch.
      \label{fig:w2}}
\end{figure}

In the second example the interaction term is given by
\begin{equation}\label{eq:interaction2}
  V(A^2)= M^4 \left(\frac{A^2}{M^2}\right)^{-1/2}, \quad \textrm{where}
  \quad M^{5}= \left(\frac{3}{8\pi G}\right)^{3/2} \, H_0^2.
\end{equation}
This corresponds  to $n=1/2$ in  equation (\ref{eq:V}).  In  this case
the triad evolution can be consistent with current observations if the
initial   value  of  $A$   is  fine-tuned.    The  initial   value  of
$B=(\dot{A}+HA)$ can be quite  arbitrary though.  Cosmic evolution for
such  a  model  and  tuned  initial conditions  is  shown  in  figures
\ref{fig:rho2}  and \ref{fig:w2}.  The  different lines  correspond to
different initial values of $B$.  Note that all of them yield the same
final  result, and  in particular  all  of them  join the  approximate
solution (\ref{eq:approximate}).   As seen in  figure \ref{fig:w2} the
equation of  state along that solution remains  constant at $w_A=-4/3$
(equation  (\ref{eq:wapproximate}))  all  the  way  until  today.   By
construction,   in  this   particular   example  $w_A\approx   -1.33$,
$dw_A/dz\approx 0$  and $\Omega_A=0.72$ today.  For  integer values of
$n$ and  fine-tuned initial conditions  it is also possible  to obtain
histories  where the  equation of  state $w_A$  remains constant  at a
value significantly smaller  than $-1$ for a long  period of time, and
just recently  approaches $-1$.  For  these models $dw_A/dz$  tends to
violate the limit in  (\ref{eq:limits}). Certain analyses of supernova
data \cite{Starobinsky} suggest that the equation of state has evolved
from  $w_{DE}\approx   0$  at   $z\approx  1$  to   $w_{DE}<-1$  today
\cite{Starobinsky}. Within the class of models (\ref{eq:V}) I have not
been able to obtain such a  behavior while keeping at the same time an
early epoch of radiation domination.

\section{Perturbations}
The  properties of  any dark  energy candidate  not only  comprise its
equation  of state,  but also  the  way their  perturbations (if  any)
behave   \cite{Wayne}.   These   perturbations  are   coupled  through
Einstein's  equations to  metric perturbations,  which in  turn affect
observables like CMB temperature  anisotropies. The evolution of triad
perturbations can also help  to determine whether models with $w_A<-1$
suffer   from  serious   instabilities   and  are   hence  unviable.   
Unfortunately, cosmological perturbation theory  with a triad turns to
be rather involved, as scalar,  vector and tensor modes couple to each
other.   Therefore, in  this section  I only  scratch the  surface and
mainly present qualitative features of the equations.

In  scalar  longitudinal gauge  and  vector  gauge,  the most  general
linearly     perturbed     spatially     flat    FRW     metric     is
\cite{Bardeen,KodamaSasaki,MFB,Wayne}
\begin{equation}\label{eq:perturbedmetric}
  ds^2=a^2(\eta)\left\{-(1+2\Phi)d\eta^2- 2 B_i\, d\eta dx^i
    +[(1-2\Psi)\,\delta_{ij}+2 h_{ij}]dx^i dx^j \right\},
\end{equation}
where $B^i$ is a transverse vector, $\partial_i B^i=0$, and $h_{ij}$
is a transverse and traceless tensor, $\partial_i h^i{}_j=h^i{}_i=0$.
In a similar way, we can decompose the perturbations of the vector
fields in the triad $\delta A^a_\mu$ into scalar and vector
components,
\begin{equation}\label{eq:decompose}
  \delta A^a_\mu=(\delta A^a_0, \partial_i\chi^a+\delta A^a_i).
\end{equation}
Here $\delta A^a_0$  and $\chi^a$ are scalars and  $\delta A^a_i$ is a
transverse  vector,  $\partial_i  \delta  A^{ai}=0$.   Note  that  for
convenience we  are using conformal  time.  In the  following, spatial
indices are raised and lowered with the metric $\delta_{ij}$.
 
\subsection{Linearized triad equations}
Because the vector fields in the triad do not couple to each other, we
can  study each  of them  at a  time.  Let  me denote  by  $A^a_i$ the
spatial components of  one of the background triad  vectors and let me
drop  in  the following  the  index  $a$.   The $0$-component  of  the
linearized equation  of motion (\ref{eq:fullmotion})  in the spacetime
(\ref{eq:perturbedmetric}) is
\begin{equation}\label{eq:constraint}
  \left(\partial_i\partial^i-2a^2\frac{dV}{dA^2}\right) \delta A_0=
-B_i A^i{}'' - \partial_i(\Phi+\Psi) A^i{}'
  +\partial_i \partial^i \chi',
\end{equation}
where a  prime denotes  a derivative with  respect to conformal  time. 
Remarkably, even  though this  is a scalar  equation, it  contains the
vector  perturbation  $B^i$.  In  that  respect, non-vanishing  vector
fields   lead    to   violations   of    the   decomposition   theorem
\cite{KodamaSasaki}.  In the absence  of a background vector quantity,
the only way to obtain a  scalar \emph{linear} in a vector $B^i$ is to
compute  its  divergence  $\partial_i   B^i$,  which  vanishes  for  a
transverse vector.  However, if the spacetime contains a non-vanishing
background vector $A_i$,  it is possible to construct  a scalar linear
in  a vector  perturbation  through the  combination  $A_i B^i$.   The
reason  why the  decomposition  theorem  is expected  to  hold in  FRW
universes  is that  in general  they  do not  have any  \emph{spatial}
direction singled out.

The $i$-components  of the linearized  equations have the form  of the
gradient of  an equation that  involves scalars plus an  equation that
involves transverse vector quantities.  The scalar equation is
\begin{equation}\label{eq:scalarequation}
  \chi''-\delta A_0'-\delta=-2a^2\frac{dV}{dA^2} \chi,
\end{equation}
where $\delta$ is implicitly defined by the decomposition
\begin{equation}\label{eq:decomposition}
  2\Phi A_i''+(\Phi+\Psi)'A_i'-2a^2\frac{d^2V}{d^2 A^2} \delta(A^2)A_i
  \equiv \partial_i \delta +\delta_i,
\end{equation} 
into a  scalar $\delta$ and  a transverse vector  $\delta_i$.  The
perturbation of $A^2$ is given by
\begin{equation}\label{eq:deltaAsquare}
  \delta(A^2)=2a^{-2}\left[A^i(\Psi \delta_{ij}
    -h_{ij})A^j+A^i(\partial_i\chi
    +\delta A_i)\right],
\end{equation}
and  in particular,  does  not contain  $\delta  A_0$. The  remaining
vector equation takes the form
\begin{eqnarray}\label{eq:vectorequation}
  \delta A_i''-\partial_j\partial^j\delta A_i
   -(A^j{}' \partial_j) B_i - 2 h_{ij}' A^j{}'-
  \delta_i=-2a^2 \frac{dV}{dA^2} \delta A_i.
\end{eqnarray}
Therefore,  again, scalar,  vector  and tensor  perturbations are  not
decoupled,  since   there  exists   a  non-vanishing  vector   in  the
background.

Note  that  (\ref{eq:constraint})  is  not a  dynamical  equation  for
$\delta A_0$, but  a constraint.  It seems that  for tachyonic models,
$dV/dA^2<0$, one cannot solve for $\delta A_0$, since in that case the
operator  $\partial_i \partial^i-2  a^2  dV/dA^2$ is  not invertible.  
This would point out to a potential inconsistency of tachyonic models.
However,  as we  shall see  next, this  difficulty is  only  apparent. 
Taking the  Laplacian of equation  (\ref{eq:scalarequation}) and using
the  time  derivative   of  the  constraint  (\ref{eq:constraint})  to
substitute  the  value  of  $\partial_i\partial^i  \delta  A_0'$,  one
obtains a first  order differential equation in time  for $\delta A_0$
that does not contain its spatial derivatives,
\begin{equation}\label{eq:constraint2}
  \frac{dV}{dA^2}\delta{A_0}'+\left[\left(\frac{dV}{dA^2}\right)'+
    2\frac{a'}{a}\frac{dV}{dA^2}\right]\delta A_0=
  f(\Phi,\Psi,\chi,a,\vec{A},\vec{B}).
\end{equation}
Here, $f$ is a function of the specified variables and its derivatives
we shall not be concerned with.  What is important is that if a set of
initial conditions satisfying  the constraint (\ref{eq:constraint}) is
specified, then, the  solution of equations (\ref{eq:constraint2}) and
(\ref{eq:scalarequation})   is    guaranteed   to   satisfy   equation
(\ref{eq:constraint}) at all times.  Hence,  it is in fact possible to
solve for $\delta A_0$.

To conclude  this part  let us count  how many ``degrees  of freedom''
(per  vector) the  triad  perturbations contain.   We  have seen  that
$\delta  A_0$  is  constrained,  so  it is  not  dynamical.   Equation
(\ref{eq:scalarequation})  contains  the  second  time  derivative  of
$\chi$ (1 dof) and  equation (\ref{eq:vectorequation}) the second time
derivative of  the transverse  vector $\delta\vec{A}$ (2  dof).  Thus,
there are  $3$ degrees in freedom  in total, as pertains  to a massive
vector.

\subsection{Perturbed  Energy Momentum Tensor} 
The  perturbations in  the triad  induce perturbations  in  its energy
momentum  tensor, which in  turn are  responsible for  sourcing metric
perturbations. In this subsection I  shall deal with vector and tensor
perturbations,  which cannot be  sourced in  conventional cosmological
models.

The perturbations  in the energy-momentum tensor of  the triad vectors
$A^a_\mu$ can  be decomposed into  an isotropic pressure  $\delta p_A$
and a  traceless anisotropic stress  $\Pi^i{}_j$ perturbation, $\delta
T^i{}_j\equiv \delta p_A \delta^i{}_j+p_A \Pi^i{}_j$.  The anisotropic
stress  itself  can  be  decomposed  into scalar,  vector  and  tensor
components,                                            $\Pi^i{}_j\equiv
{}^{(s)}\Pi^i{}_j+{}^{(v)}\Pi^i{}_j+{}^{(v)}\Pi_j{}^i+{}^{(t)}\Pi^i{}_j$,
where                 $^{(s)}\Pi^i{}_j=(\partial^i\partial_j-\delta^i_j
\partial^k\partial_k/3)\Pi$,     $^{(v)}\Pi^i{}_j=\partial_j    \Pi^i$
($\Pi^i$  transverse),  and   ${}^{(t)}\Pi^i{}_j$  is  transverse  and
traceless.
 
The  equations  of  motion   for  vector  (metric)  perturbations  are
\cite{Wayne,KodamaSasaki}
\begin{equation}\label{eq:vectormetric}
  \partial_j B^i{}'+2 \frac{a'}{a}\partial_j B^i{}
  =8\pi G a^2 p_A {}^{(v)}\Pi^i{}_j,
\end{equation}
where $^{(v)}\Pi^i{}_j$  is the vector  part of the  triad anisotropic
stress tensor.  In the absence of anisotropic stress sources, equation
(\ref{eq:vectormetric}) implies that  vector perturbations decay away,
$B^i \propto 1/a^2$. So even if they are generated in the early stages
of the universe,  they are not expected to  be significant today.  The
transverse   and   traceless   anisotropic   stress   sources   tensor
perturbations, i.e. gravitational waves,
\begin{equation}
  h_{ij}{}''+2\frac{a'}{a}h_{ij}{}'-\partial^k\partial_k h_{ij}=
  8\pi G p_A {}^{(t)}\Pi_{ij}.
\end{equation}
As  opposed to  vector perturbations,  in the  absence of  sources the
amplitude  of long-wavelength  gravitational waves  remains  constant. 
Hence, if  they are primordially produced, say  during an inflationary
stage, they could still have a sizable amplitude today.

I  shall not  write  down all  the  terms that  the perturbed  spatial
components  of the  energy momentum  tensor of  a single  triad vector
contains.   They  straightforwardly (but  tediously)  follow from  the
insertion   of   the   perturbations  (\ref{eq:perturbedmetric})   and
(\ref{eq:decompose}) into  equation (\ref{eq:EMT}).  Instead,  for the
sake of illustration I shall consider only
\begin{equation}
  \delta[{}^{a}T^i{}_j]= 2 a^{-2} \frac{dV}{dA^2}
  \left(A^i \delta A_j+\delta A^i A_j\right)+\cdots,
\end{equation}
where the dots  denote the multiple terms I  am not explicitly writing
down.   In  order  to  study   the  evolution  of  vector  and  tensor
perturbations we have  to compute the vector and  tensor components of
the  previous  expression.   In  Fourier  space, these  are  given  by
\cite{KodamaSasaki}
\begin{equation}
 p_A{}^{(v)}\Pi^i=-\sum_a\frac{2i a^{-2}}{k^2}\frac{dV}{dA^2}
 (\vec{k}\cdot\vec{A}^a)\, \delta A^{ai},
\end{equation}
\begin{eqnarray}\label{eq:tt}
 p_A {}^{(t)}\Pi^i{}_j=2a^{-2}\frac{dV}{dA^2}\sum_a
   \Bigg[A^{ai} \delta A^a_j+\delta A^{ai} A^a_j&-&
   \left(\frac{k^i}{k}\delta A^a_j+\delta A^{ai}\frac{k_j}{k}\right)
   \frac{\vec{k}\cdot\vec{A}^a}{k}- \nonumber \\ 
   &&{}-\left(\delta^i_j-\frac{k^i k_j}{k^2}\right)\vec{A}^a\cdot
   \delta\vec{A}^a\Bigg],
\end{eqnarray}
which as required are transverse and traceless.  Note that we sum over
the three  triad vectors  to obtain the  total energy  momentum tensor
perturbation.  Because the  triad perturbations $\delta{A}^{ai}$ are a
priori  totally independent  from each  other, the  anisotropic stress
${}^{(t)}\Pi^i{}_j$ does  not vanish in general.   Thus, vector fields
are not only expected to  source vector perturbations, but also tensor
perturbations.  Again, the reason is that the decomposition theorem is
violated. With  the aid  of the background  vectors it is  possible do
construct   traceless  and   transverse  quantities   linear   in  the
perturbations.

If  the perturbations  have certain  symmetries  , ${}^{(t)}\Pi^i{}_j$
does indeed  vanish (for  the particular term  in the  energy momentum
tensor we are  considering).  Because $A^{ai}$ and $k^i$  are the only
vectors  in the  problem, the  triad perturbation  $\delta{A}^{ai}$ is
expected to be a function  of $A^{ai}$ and $k^i$.  Since $\delta{A}^a$
is transverse, it has to be of the form
\begin{equation}\label{eq:transverse}
  \delta{A}^{ai}=\alpha\cdot 
  \left(\delta^i{}_j -\frac{k^i}{k}\frac{k_j}{k}\right) A^{aj}+
  \beta\cdot \varepsilon^i{}_{jk}A^{aj}k^k,
\end{equation}
where   $\alpha$   and   $\beta$   are  two   scalar   functions   and
$\varepsilon_{ijk}$  is totally  antisymmetric.  Assume  that $\alpha$
and  $\beta$ do not  depend on  the index  $a$ (that  is, they  do not
depend    on    $k_i    A^{ai}$.)    Then,    substituting    equation
(\ref{eq:transverse}) in (\ref{eq:tt}) and using (\ref{eq:ansatz}) one
finds not only that ${}^{(t)}\Pi^i{}_j=0$, but also ${}^{(v)}\Pi^i=0$.

\subsection{Stability}
The main worry  one faces when dealing with  tachyonic fields is their
stability.  In  order to  figure out to  what extent the  solutions we
have found in Section  \ref{sec:evolution} are stable, we should solve
the  system of  cosmological  perturbation equations  just presented.  
Obviously,   the   coupling   between   scalar,  vector   and   tensor
perturbations makes  this task quite  formidable.  In this  section we
dramatically simplify the equations by neglecting metric perturbations
and concentrating  on a particular set  of vector modes.   The hope is
that this drastic simplification  captures the qualitative features of
the equations.

So   let's  set  $\Phi=\Psi=B^i=h_{ij}=0$   and  consider   the  triad
perturbation    equations   in    Fourier   space.     From   equation
(\ref{eq:decomposition}),   for    modes   for   which   $\vec{k}\cdot
\vec{A}=0$, it follows that  $\delta=0$.  Then, $\chi=\delta A_0=0$ is
a     solution      of     equations     (\ref{eq:constraint})     and
(\ref{eq:scalarequation}).  Because $\delta_i\propto (A^j\delta A_j)\,
A_i$,   if   in  addition   we   consider   perturbations  such   that
$\delta\vec{A}\cdot   \vec{A}=0$,   the   remaining  vector   equation
(\ref{eq:vectorequation}) reads
\begin{equation}\label{eq:stabilityequation}
  \delta\ddot{A}_i+H \delta \dot{A}_i+
  \frac{k^2}{a^2}\delta A_i+2\frac{dV}{dA^2}\delta A_i=0,
\end{equation}
where  for convenience  I  have gone  back  to cosmic  time. Note  how
$2dV/dA^2$ plays the  role of a mass term in  the last equation.  This
is   why  I   call  interactions   with  $dV/dA^2<0$   ``tachyonic''.  
Generically, if $dV/dA^2$ is negative, one expects growing modes, that
is, instabilities.

In order to check the stability  of the triad, it suffices to consider
long-wavelength        modes,         $k=0$,        in        equation
(\ref{eq:stabilityequation}). For sufficiently  high $k$, the gradient
dominates the interaction and solutions are stable. Hence, any form of
instability is an ``infrared'' effect, rather than an ultraviolet one.
For  a given  expansion,  equation (\ref{eq:a}),  along the  attractor
(\ref{eq:mattersolution}),  the long-wavelength  solution  of equation
(\ref{eq:stabilityequation}) is
\begin{equation}\label{eq:perturbationsolution}
  \frac{\delta A_i}{a}=C_+\,t^{\gamma_+}+C_-\, t^{\gamma_-},
  \quad \text{where}\quad \gamma_+=\frac{1}{n+1}\quad \text{and} \quad
  \gamma_-=-3\beta+\frac{n}{n+1}.
\end{equation}
Here,  $C_+$   and  $C_-$  are  two  integration   constants  and  for
convenience I have divided by the scale factor to obtain the length of
the perturbation.  We should compare these solutions with the behavior
of  the background,  equation (\ref{eq:mattersolution}).   Recall than
$A$ is the  common length of the background  triad vectors.  The $C_+$
mode in equation (\ref{eq:perturbationsolution})  grows as fast as $A$
and   the    $C_-$   mode   grows    less   rapidly   than    $A$   if
$n(1-3\beta)<1+3\beta$.  Therefore, the  $C_-$ mode decays relative to
$A$ during  inflation, radiation  and dust domination.   Hence, within
the scope of  our analysis, the system is  not unstable throughout the
period where the triad is subdominant\footnote{Strictly speaking it is
  not stable  either, as  the relative perturbations  do not  decay.}. 
Note however  that the triad would  be unstable for  certain values of
$n$  if  there  had been  a  period  of  cosmic history  during  which
$\beta<1/3$.

When the  triad becomes  dominant, the vector  evolves towards  the de
Sitter   attractor    (\ref{eq:dS}).   The   solution    of   equation
(\ref{eq:stabilityequation}) along the de Sitter attractor is
\begin{eqnarray}
  \frac{\delta A_i}{a}=C_+ \exp (H_* t)+C_- \exp (-2H_* t),
\end{eqnarray}
where $H_*$  is the value of  the Hubble constant along  the de Sitter
solution.  Note  that along the  latter $A$ itself is  constant. Thus,
the de Sitter attractor is  unstable, in the sense that $\delta A_i/a$
(the length  of the  vector perturbation) grows  relative to  $A$ (the
length of the background  vector).  Because during the previous stages
of  cosmic  history (when  the  triad  was  subdominant) the  relative
amplitude of  the perturbations  $\delta A/A$ has  remained unaltered,
the time  perturbations become relevant will depend  on early universe
initial  conditions.  If  the primordial  vector amplitude  agrees for
instance with the (scalar)  amplitude of density fluctuations, $\delta
A/A \approx  10^{-5}$, $\delta  A/A$ becomes of  order one  about $12$
$e$-folds after the  onset of triad domination.  By  then the universe
has presumably grown anisotropic, because there is no reason to expect
that  perturbations   in  the   three  different  triad   vectors  are
correlated.   We'll  have to  wait  several  billion  years till  that
happens.  At  present the  effect is still  small, since  the relative
amplitude has increased at most by a factor of order one.  In fact, it
is tempting to speculate whether some of the anomalies observed in the
CMB  radiation  \cite{quadrupole},  specially  in the  quadrupole  and
octopole  (see   however  \cite{low?})   and  the   related  hints  of
statistical anisotropy  \cite{non-isotropy}, might  be due to  such an
instability, which sets in once  the universe starts to accelerate and
mostly affects large scales.

\section{Summary and Conclusions}
In  this paper  I  have considered  whether  a vector  field could  be
responsible  for the current  stage of  cosmic accelerated  expansion. 
The existence of a  vector with non-vanishing spatial components turns
to  be compatible  with the  isotropy of  a Friedmann-Robertson-Walker
universe provided  the vector is part  of a ``cosmic  triad'', i.e.  a
set of  three identical vector fields pointing  in mutually orthogonal
directions.   A  set   of  three  identical  self-interacting  vectors
naturally  arises for  instance  in  a gauge  theory  with $SU(2)$  or
$SO(3)$ gauge group.

A distinctive property of a cosmic triad is that its equation of state
of can become  less than $-1$, even though its  kinetic terms have the
conventional   form.    The    necessary   condition   is   that   the
self-interaction is  ``tachyonic'', i.e.  it  naively gives rise  to a
negative squared vector mass.  Although the simple analogy with scalar
tachyons suggests  that the study  of tachyonic vectors  is justified,
there  are also arguments  that connect  tachyonic vectors  to phantom
particles \cite{Dubovsky}.  In analogy to tracking quintessence models
\cite{tracker}, in  this paper I  have explored a particular  class of
tachyonic models, where the  interaction is an inverse power-law.  For
appropriate   initial  conditions,   there   exist  seemingly   viable
cosmologies  where  dark  energy  has  an  equation  of  state  $w<-1$
throughout  cosmic history, including  today.  In  top of  that, these
models have attractors that  render cosmic evolution quite insensitive
to initial conditions and  yield equations of state sufficiently close
to $-1$ today.  The experimental  constraints on the value of the dark
energy   equation  of  state   restrict  the   power  of   the  vector
self-interaction, while the time  at which cosmic acceleration sets in
determines  its energy  scale.   Within  the class  of  models I  have
considered, the current limits on the equation of state of dark energy
require the  vector self-interaction to be a  non-analytic function of
its squared length.

Finally,  I have  scratched the  surface of  cosmological perturbation
theory in the presence of  a cosmic triad. The most remarkable feature
is the violation  of the decomposition theorem.  In  the presence of a
vector which has non-vanishing  spatial components, scalar, vector and
tensor perturbations do not  decouple from each other.  In particular,
metric vector  perturbations show up  in scalar equations, and  in the
absence of particular symmetries in the perturbations vectors are able
to source tensors.  The (scalar)  time components of the triad vectors
are not  dynamical, i.e.  they  are constrained.  Despite  an apparent
difficulty, it is possible to  solve the constraint also for tachyonic
models.   I  have  also   considered  solutions  of  the  perturbation
equations for  a particular  set of modes,  under the  assumption that
metric perturbations  are negligible. During  inflation, radiation and
dust  domination  the  relative  perturbations  in  the  triad  remain
constant.  However, there is  a long wavelength instability during the
late-time stage  of de Sitter acceleration,  where triad perturbations
grow relative to the background.  Hence, the time the universe becomes
anisotropic  depends  on   early  universe  initial  conditions.   For
reasonable  primordial   perturbation  amplitudes,  the   universe  is
expected to  become anisotropic  long time after  the onset  of cosmic
acceleration.   The instability of  the triad  during this  epoch also
suggests a possible relation between  the large angle anomalies in the
CMB  sky and the  onset of  cosmic acceleration,  but further  work is
needed to test this idea.  A more careful investigation is also needed
to establish whether  tachyonic vectors are fully stable,  and how the
inclusion of  metric perturbations affects  the behavior of  the triad
perturbations (and vice versa).

To conclude, at the level of the present analysis it seems that vector
fields could indeed be responsible  for the present stage of late time
cosmic acceleration,  though it is  yet unclear how  quantum mechanics
constrains the tachyonic models I have studied here.

\begin{acknowledgments}
  This paper is  dedicated to Isabel.  It is a  pleasure to thank Sean
  Carroll, Sergei Dubovsky, Vikram  Duvvuri, Wayne Hu, Lam Hui, Eugene
  Lim, Geraldine  Servant and Paul Steinhardt  for useful discussions,
  and the KICP  for software assistance. This work  has been supported
  by the US DOE grant DE-FG02-90ER40560.
\end{acknowledgments}

\appendix*

\section{A Non-Abelian Triad Realization} 
In  this  appendix, I  show  that  a cosmic  triad  can  arise from  a
non-Abelian  $SU(2)$  gauge theory.  For  that  purpose, consider  the
gauge-invariant Yang-Mills Lagrangian
\begin{equation}\label{eq:YM}
  -\frac{1}{4}\int d^4x \sqrt{-g}\, F^a_{\mu\nu} F_a^{\mu\nu},
\end{equation} 
where $F^a_{\mu\nu}$ is the non-Abelian field-strength
\begin{equation}
  F^a_{\mu\nu}=\partial_\mu A_\nu^a-\partial_\nu A^a_\mu+
  \varepsilon^a{}_{bc}A^b_\mu A^c_\nu.
\end{equation}
The  totally antisymmetric  tensor $\varepsilon^a{}_{bc}$  encodes the
structure  constants of  the  $SU(2)$ Lie-algebra.   The equations  of
motion of the fields in an arbitrary spacetime are
\begin{equation}
  \frac{1}{\sqrt{-g}}\partial_\mu\left(\sqrt{-g}F^a{}^{\mu\nu}\right)
  +\varepsilon^a{}_{bc}A^b{}_\mu F^{c\mu\nu}=0.
\end{equation}
The ansatz  (\ref{eq:ansatz}) satisfies  these equations of  motion in
the FRW universe (\ref{eq:FRW}) if
\begin{equation}\label{eq:namotion}
\ddot{A}+3H\dot{A}+\left(H^2+\frac{\ddot{a}}{a}\right)A+
\frac{dV}{dA}=0,
\end{equation}
where
\begin{equation}\label{eq:naV}
  V=\frac{1}{2}\left(A_\mu A^\mu\right)^2.
\end{equation}
Equation (\ref{eq:namotion}) agrees with  the triad equation of motion
(\ref{eq:xmotion}).  The energy momentum tensor of the gauge fields is
\begin{equation}
  T_{\mu\nu}=F^a_{\mu\rho}F_a{}_\nu{}^\rho-
\frac{1}{4} F^a_{\rho\sigma} F_a^{\rho\sigma} g_{\mu\nu}.
\end{equation}
It can be verified that $T^0{}_i=0$, whereas
\begin{equation}
  -T^0{}_0=\frac{3}{2}(\dot{A}+H A)^2+3V,\quad
  T^i{}_j=
  \left(\frac{1}{2}(\dot{A}+H A)^2-3V+2\frac{dV}{dA^2} 
    A^2\right)\delta^i{}_j.
\end{equation}
The self-interaction  $V$ is again  given by equation  (\ref{eq:naV}). 
Therefore, the  energy density and  pressure of the  non-Abelian gauge
fields agree with the ones of the triad, equations (\ref{eq:pandrho}).
This equivalence between the triad and the non-Abelian gauge fields in
the symmetric case we are considering is confirmed by substituting the
ansatz  (\ref{eq:ansatz})  into   the  actions  (\ref{eq:action})  and
(\ref{eq:YM}). Note that  for the triad vectors, as  opposed to scalar
fields or perfect  fluids, the Lagrangian density is  not the pressure
nor  the  energy  density.   The  existence  and  some  properties  of
non-trivial  solutions of  general  relativity coupled  to an  $SU(2)$
gauge   field   in   a   FRW   universe  have   been   considered   in
\cite{non-abelian}.

\end{document}